\documentclass[11pt]{article}

\usepackage{amsbsy}
\usepackage{amsmath}
\usepackage{amssymb}
\usepackage{graphicx}
\usepackage{graphics}
\usepackage{latexsym}
\usepackage{cite}

\begin{document}
%

%%%%%%%%%%%%%%%%%%%%%%%%%%%%%%%%%%%%%%%%%%%%%%%%%%%%%%%%%%%%%%%%%%
%
%
%
%                Definitions
%
%
%
%%%%%%%%%%%%%%%%%%%%%%%%%%%%%%%%%%%%%%%%%%%%%%%%%%%%%%%%%%%%%%%%%%
%

\newcommand{\sect}[1]{\setcounter{equation}{0}\section{#1}}
\renewcommand{\theequation}{\thesection.\arabic{equation}}
\setcounter{equation}{0}

\newcommand{\di}[2]{\frac{d {#1}}{d {#2}}}
\newcommand{\de}[2]{\frac{\partial {#1}}{\partial {#2}}}

\def\spazio#1{\vrule height#1em width0em depth#1em}
\def\ad{a^\dagger}
\def\bra#1{\langle#1|}
\def\ket#1{|#1\rangle}
\def\scal#1#2{\langle#1|#2\rangle}
\def\matrel#1#2#3{\langle#1|#2|#3\rangle}
\def\pert#1{|\psi_n^{(#1)}\rangle}
\def\scalpert#1#2{\langle#1|\psi_n^{(#2)}\rangle}
\def\thm#1#2 { \textsc{Theorem #1.#2}\phantom{X}}
\def\rmk#1#2 { \textsc{Remark #1.#2 }\phantom{X}}
\def\dim { \textsc{Proof.} \phantom{X}}
\def\spazio#1{\vrule height#1em width0em depth#1em}
\def\enr{E_{{}_{\mathrm{NR}}}}

\setlength{\unitlength}{1cm}
\newcommand{\be}{\begin{eqnarray}}
\newcommand{\ee}{\end{eqnarray}}
\newcommand{\bee}{\begin{eqnarray*}}
\newcommand{\eee}{\end{eqnarray*}}
\newcommand{\ra}{\rightarrow}
\newcommand{\sse}{\subsection}
\newcommand{\+} {\`}
\newcommand{\hb} {\hbar}
\newcommand{\R}{\mbox {\sc R}}
\newcommand{\N}{\mbox {\sc N}}
\newcommand{\Z}{\mbox {\sc Z}}
\newcommand{\C}{\mbox {\sc C}}
\newcommand{\D}{\mbox {\sc D}}
\newcommand{\I}{\mbox {\sc 1}}
\newcommand{\T}{\mbox {\sc T}}
\newcommand{\s}{\mbox {\sc P}}
\newcommand{\p}{\mbox {\sc T}}
\newcommand{\0}{\mbox {\sc 0}}
\newcommand{\Rp}{\mbox {\sc r}}
\newcommand{\Np}{\mbox {\sc n}}
\newcommand{\Zp}{\mbox {\sc z}}
\newcommand{\Cp}{\mbox {\sc c}}

%
%
%%%%%%%%%%%%%%%%%%%%%%%%%%%%%%%%%%%%%%%%%%%%%%%%%%%%%%%%%%%%%%%%%%
%

\baselineskip= 20 pt
{\begin{center}
{\LARGE{\bf PT-symmetric operators and metastable states  of the 1D relativistic oscillators.} }          \end{center}}
\baselineskip= 14 pt

\bigskip
\centerline{Riccardo Giachetti}
\medskip
\centerline{{  Department of Physics, University of
Firenze and I.N.F.N. Sezione di Firenze }}\centerline{{ Via G.
Sansone 1, 50019 Sesto Fiorentino, Firenze, Italy\footnote{
e-address:giachetti@fi.infn.it} }}
\bigskip

\centerline{Vincenzo Grecchi}
\medskip
\centerline{{  Department of Mathematics, University of
Bologna and I.N.F.N. Sezione di Bologna }}\centerline{{ Piazza di Porta S. Donato 5,
40126 Bologna, Italy\footnote{
e-address:grecchi@dm.unibo.it} }}
\bigskip\medskip

\begin{abstract}

We consider    the one-dimensional Dirac equation for the harmonic oscillator and the associated second order separated operators  giving the resonances of the problem by complex dilation. The same operators have  unique extensions as  closed \textit{PT}-symmetric operators defining  infinite positive energy levels converging to the Schr\"odinger ones as $c\ra\! \infty.$ Such energy levels and their eigenfunctions give directly a definite choice of metastable states of the problem.
%Moreover, they are the energy levels and the stationary states of a new relativistic model similar to the Dirac one.
Precise numerical computations shows that these levels  coincide with the  positions of the resonances up to the  order of the width.
Similar   results are found for the Klein-Gordon oscillators, and in this case there is an infinite number of dynamics and the eigenvalues and eigenvectors of the \textit{PT}-symmetric operators give metastable states for each dynamics.

\hfill \textit{\textbf{ }}
\end{abstract}

\bigskip

\sect{Introduction.}
\label{intro}

The harmonic oscillator is one of the fundamental
dynamical systems both in classical and quantum theory
\cite{MoSmi} and the natural relativistic extensions
have been investigated since a long time.
In the relativistic quantum mechanics literature,
therefore,  one can find many different models
that are considered relativistic oscillators
\cite{ZBJ,CVH,VD,Ruij,ToNo,MoSmi,Tha}.
These share the common property of reducing to
the usual quantum oscillator in the Schr\"odinger
limit but, otherwise, they present large differences
in the way the interaction is constructed.
Obviously different models have different dynamical and
spectral properties: for instance scalar potentials have
been added to vector potentials in a phenomenological
way with the purpose of stabilizing the system and giving
confinement, \cite{FKR,ABA,CLM,Martin}, or else the
interaction has been obtained by adding to the free Dirac
Hamiltonian a linear term in the position
coordinate twisted by the Dirac $\beta$ matrix
\cite{MosSz,Cas,Nou,LR,MV,MoNi,Str}. The relevance of such
models in physics is clear and goes beyond their
use for an elementary description of the confinement
mechanism, in view of the connection with QFT and of the
application of the Dirac equations to new materials
\cite{Kats}, where the relativistic effects are more easily
measured.

We here call spinor and scalar
relativistic oscillator, respectively, the Dirac and the
Klein-Gordon equations with a quadratic electrostatic
potential because these models are very realistic and, in
principle, can be experimentally implemented. Although they
do not satisfy the necessary conditions for the
exactness of the Foldy-Wouthuysen transformations
\cite{Nik,BG09}, it is obvious that the well known harmonic oscillator
is reproduced in the Schr\"odinger limit. Despite
their apparent simplicity, however, the study
of the relativistic electrostatic oscillators
is not so easy as it looks at first glance,
but the results are  satisfying.

Common wisdom tells us that strong instability
is induced in quantum relativistic equations with
unbounded potentials in the way described by the Klein
paradox, \cite{Kre}. Because of such instability
the strongly bounded relativistic systems should
present continuous spectrum and complex resonances.
This picture is well signified by a result that
Plesset established in 1932  \cite{Plesset} and that can be
summarized, with a rather paradoxical formulation,
as follows: the Dirac equation with a minimal coupling
that involves an electrostatic ``bounding'' potential
does not admit bound states. Of course, since
bound states are present in the Schr\"odinger
limit, the natural question that arises is what
happens in the transition from a non-relativistic
to a relativistic regime and in which sense this
transition can be considered continuous in $1/c$.
A first answer was given by Titchmarsh in \cite{Tit}
who studied in detail the perturbation treatment of the
Dirac equation with piecewise linear vector potential.
Due to the very cumbersome analytical calculations involved,
the treatment was essentially restricted to the
first perturbation order giving, however some more
explicit informations on the properties of the continuous
spectrum by investigating the Weyl function $m(\lambda)$
of the singular boundary value problem \cite {CL}
in the complex plane.
Very recently numerical investigations of the Dirac equation
with a linear and a quadratic electrostatic potential
were presented in \cite{GS,GG}, looking for a dissipative
model: metastable states were found and
the Schwinger pair production rate \cite{Sch}
was calculated in terms of the spectral concentration
and in terms of the imaginary part of the resonances of
the Dirac equation in external linear and quadratic
potentials. In \cite{GS} the density of the states
\cite{K} was determined at finite values of $(1/c)$,
finding a sum of Breit-Wigner lines whose width reproduced
the pair production rate. In \cite{GG} the Dirac equation
with a quadratic potential was revisited in the framework of
the large order perturbation theory. The singular
problem gave rise to asymptotic non oscillating series and the
sum was calculated by the Distributional Borel (DB)
method, \cite{CGM1,CGM2,CGM3,CGM4,CGM5}, coupled with a Pad\'e
approximation for the DB transform, finding complete
agreement with \cite{GS}. Moreover the spectral
problem was clearly formulated stressing the reasons
for the dissipative choice.
It was also shown that this choice leads to the study
of non self-adjoint operators defined by complex dilation:
the imaginary part of the complex energy levels, interpreted
as resonances, again reproduces the pair production.

Although exact bound states do not exist, in this paper we
show that particular metastable states can
be identified. They are uniquely obtained from eigenvalues of
the decoupled equations of Titchmarsh.
At least for a small ratio of the interaction
to the mass energy they are well approximated.
The role of the corresponding levels is similar
to the role of the real part of the resonances.
In this case we have both and it is our
purpose to compare them. Actually, for the metastable states we have more exact results. In fact the
Klein-Gordon like Hamiltonians,
obtained by Titchmarsh in \cite{Tit} (see also
\cite{LL}) from the separation of the
Dirac system of equations -- and hereby referred
as to Titchmarsh Hamiltonians --
are uniquely defined closed operators with a discrete spectrum
that gives the metastable levels of the Dirac equation.
The Titchmarsh Hamiltonians exhibit what in current literature
is known as the \textit{PT}-symmetry, where $P$ is
the parity transformation and $T$ the complex conjugation
\cite{BG,ZBJ,BB}: our paper, therefore, adds also a
meaningful contribution to the discussion on
the physical meaning of the $PT$-symmetric operators that still
keep arising great interest
(see, for instance, \cite{AM1,AM2,GuSa,BBCW,BCMS}).
The metastable  levels are studied in the next two sections.
In section 2 we use the methods of the functional analysis
supported by some recent results coming from the
study of anharmonic quantum oscillators \cite{BG,SH}
and we give formal statements of our analytical results.
We also show the possible role of the metastable levels as
stationary levels of the separated \textit{PT}-symmetric dynamics.
In section 3 we then present the complex extensions of the
position variables we have used,
that have been introduced for two reasons: a better variational
approximation and the connection with known results, mainly
those of references \cite{BG,SH}.
In the final section we present a discussion
of the numerical results connecting the metastable levels
to the resonances of the model.
We have thus calculated some perfectly defined energy levels
of the Titchmarsh Hamiltonians stable at the Schr\"odinger limit and
we have established their relationship to the real part of the resonances
determined  in \cite{GG} by the the DB sum: we have thus proved that the difference
is of the second order in the imaginary part of the
resonances  themselves and it can therefore be ascribed
to the pair production rate, as it had to be
expected on a physical ground.
We have found that the best numerical way,
reaching the very high precision necessary in order to compare
the asymptotic behavior in a parameter
$\Omega \sim O(1/c^2)\rightarrow 0$ with the DB sum of the
power series expansion  \cite{GG} is provided by a
specialization of the Rayleigh-Ritz scheme \cite{WA} obtained through the matrix
moments method \cite{GG_Matrix,BF}. This method could also be used for higher values of $\Omega$ where,
however, other approaches could be equally or even more efficient
\cite{GS,MoGroups}

\bigskip

\sect{The Dirac and the Klein-Gordon one-dimensional oscillators.}
\label{oscillators}

Let us consider the one-dimensional Dirac equation in an electrostatic potential $V(x)$ \cite{LL,Tit}.
Using the notations of ref. \cite{Tit} we assume a two-component spinor wave function of the form
$X={}^t[X_2(x),X_1(x)]$ so that the explicit form of the Dirac equation reads
\begin{eqnarray}
\frac 1c\bigl(W+mc^2- V(x)\bigr)X_1(x)-\hbar\,\frac{d}{dx}\,X_2(x)=0\cr\spazio{1.8}
\hbar\,\frac{d}{dx}\,X_1(x)+\frac 1c\bigl(W-mc^2-V(x)\bigr)X_2(x)=0
\label{DiracEquation}
\end{eqnarray}
We therefore see that the `large' component of the spinor is $X_2(x)$.
In the following we will assume
$$V(x)=\frac 12\,m\omega^2x^2. $$

We rescale the spatial coordinate as
$$x\mapsto \bigl(\frac{m\omega}\hbar\bigr)^{1/2}x$$
so to make it dimensionless and we define the equally
dimensionless parameters
$$\Omega=\bigl(\frac{\hbar\omega}{4mc^2}\bigr), \qquad E=\frac 2{\hbar\omega}\,(W-mc^2)\,.$$
From (\ref{DiracEquation}) we then get the system
\begin{eqnarray}
&{}&\frac{d}{dx}\,X_2(x)-\sqrt{\Omega}\,\bigl( E-x^2+\frac 1{\Omega}\bigr)\,X_1(x)=0\spazio{1.0}\cr
&{}&\frac{d}{dx}\,X_1(x)+\sqrt{\Omega}\,\bigl( E-x^2\bigr)\,X_2(x)=0
\label{DiracAdim}
\end{eqnarray}

In order to reduce the system (\ref{DiracAdim}) to separate second order equations, it is convenient to introduce the linear combinations
\begin{eqnarray}
\psi_+(x)=\frac 1{\sqrt{2}}\,\bigl(X_1(x)+ iX_2(x)\bigr)\,\qquad \psi_-(x)=-\frac {i}{\sqrt{2}}\,\bigl(X_1(x)- iX_2(x)\bigr),
\label{psipm}
\end{eqnarray}
%with inverse
%$$X_1(x)=\frac 1{\sqrt{2}}\,\bigl(\psi_+(x)+i\psi_-(x)\bigr)\,\qquad X_2(x)=-\frac {i}{\sqrt{2}}\,\bigl(\psi_+(x)-i\psi_-(x)\bigr),$$
yielding
\begin{eqnarray}
&{}&\frac{d}{dx}\psi_+(x)-i\sqrt{\Omega}\,\bigl(E-x^2)\psi_+(x)-\frac {i}{2\sqrt{\Omega}}\,\bigl(\psi_+(x)+i\psi_-(x)\bigr)=0\spazio{1.0}\cr
&{}&\frac{d}{dx}\psi_-(x)+i\sqrt{\Omega}\,\bigl(E-x^2)\psi_-(x)+\frac 1{2\sqrt{\Omega}}\,\bigl(\psi_+(x)+i\psi_-(x)\bigr)=0\,.
\label{Diracpsi}
\end{eqnarray}
Solving the first equation (\ref{Diracpsi}) in $\psi_-(x)$,
$$ \psi_-(x)=-2\sqrt{\Omega}\,\frac{d}{dx}\psi_+(x)+i\bigl(\,2\Omega(E-x^2)+1\,
\bigr)\,\psi_+(x)$$
and substituting in the second equation, we finally find for $\psi_+(x)$
the second order equation
\begin{eqnarray}
\widetilde {H}_+({{\Omega}},E)\psi_+(x)=\lambda\psi_+(x)
\nonumber
\end{eqnarray}
where $\lambda=\lambda(\Omega,E)=E+{\Omega}E^2$ and where the Titchmarsh Hamiltonian
\begin{eqnarray}
\widetilde {H}_+({{\Omega}},E)\,=
-\frac{d^2}{dx^2}- 2i{\sqrt{\Omega}} x
+(1+2E{\Omega})x^2-\Omega x^4
\label{OFKG}
\end{eqnarray}
is a formal operator in the Hilbert space of the states $\mathcal{H}=L^2(\mathbb{R})$. Here and in the following we will always take $\sqrt{\Omega}>0$.

The assumption of a real energy $E$ will
prove to be self-consistent with the boundary value problem that turns
out to be complete, in the sense that the formal operator $\widetilde {H}_+({{\Omega}},E)$
extends uniquely to a closed Hamiltonian $H_+={H}_+({{\Omega}},E)$ (with discrete spectrum).
The unstable term $-{\Omega}x^4$ is still present in $H_+$, suggesting that
the Klein paradox arguments could be brought to bear: the imaginary term
$- 2i{\sqrt{\Omega}} x$, however, makes the
problem complete and the {\it{PT}}-symmetric extension of
$H_+$ unique. Moreover the unstable quartic term can be easily dealt with. Indeed in \cite{BG} a direct demonstration was given of the equality of the eigenvalues of a pair of quantum systems: the double well and the unstable anharmonic oscillator defined by complex translation, both of them, obviously, making sense and being well defined. In \cite{SH}, then, a general proof was presented that the spectrum of $H_+$ is
discrete and positive:
$$\sigma(H_+)=\{\lambda_n>0\}_{n\in{\mathbb{N}}}\,.$$
The equivalent adjoint Hamiltonian  is
\begin{eqnarray}
H_-=PH_+P=TH_+T=H_+^*\,,
\label{H_minus}
\end{eqnarray}
where $P$ is the parity transform and $T$ is the time reversal.

Let us also notice that the non self-adjointness of the Hamiltonians has a physical meaning: it allows to define uniquely and to distinguish the $H_\pm$ operators.
Finally, the labeling of the levels is the same we have at the Schr\"odinger limit, in the
hypothesis that only one level is associated to one eigenvalue (see equation (\ref{RELAM}) here below). We shall argue that this is the case and, moreover,
we expect and we prove the existence of infinite positive levels, stable at ${{\Omega}}=0$.
Let us now express our results in a formal way.
$\spazio{0.8}$

\thm 21 \textit{Let $\Omega>0,$ $E\in\mathbb{R}$, then it is uniquely defined the closed $PT$-symmetric extension with discrete spectrum,
${H}_+({{\Omega}},E)$ of the operator $\tilde{H}_+(\Omega, E)$. The closed operator $H_-(\Omega, E)$
is then defined from $(\ref{H_minus})$.}

\dim We uniquely define the operator $H_+$ by the $L^2$ conditions at $\pm\infty.$
The asymptotic behavior of the fundamental solution for $x\ra+\infty$ is
\begin{eqnarray}
\Psi_+(x)\sim \frac1{x}\,\exp\bigl({iS(x) -\ln(x)}\bigr) =
\frac1{x^2}\,\exp\bigl({iS(x)}\bigr)\,,
\nonumber
\end{eqnarray}
where
\begin{eqnarray}
 S(x)=\int^x\sqrt{\Omega y^4-(1+2E\Omega)y^2}\,\,dy \sim \sqrt{\Omega }\,\frac{x^3}{3}-\frac{(1+2E\Omega)}{2\sqrt{\Omega}}x,
\label{Action}
\end{eqnarray}
the other having the behavior
\begin{eqnarray}
~~~~~
\Psi_-(x)\sim \frac1{x^2}\,\exp\bigl({iS(x)}\bigr)\quad {\mathrm{as}} \quad x\ra-\infty\,.
~~~~~ \Box
\nonumber
\end{eqnarray}

\thm 22 \textit{Let $E\in\mathbb{R},$ $\Omega>0,$ and fix the index state $\pm$, there are infinite positive simple eigenvalues $\lambda_n(\Omega,E)$  $n\in\mathbb{N}$ of the Titchmarsh operators for the harmonic oscillator.  For each $\Omega>0,$ there are infinite positive energy levels  $E_n(\Omega)$ of the spectrum of the Titchmarsh operators.}

\dim Let us fix $\Omega>0,$ $n\in{\mathbb{N}}$ and recall that, according to \cite{SH}, for any $E\in\mathbb{R}$ we have positive eigenvalues $\lambda_n(E)$ of $H_\pm(E).$
We thus signify by ``level'' any generalized eigenvalue $E_n>0,$
solution of the implicit equation ${\Omega} E^2+E=\lambda_n(E)$, namely
\begin{eqnarray}
E=\frac 1{2\Omega}\, {\bigl(\sqrt{1+4\lambda_n(E)\Omega}-1\bigr)}
\label{RELAM}
\end{eqnarray}
The existence of a solution of (\ref{RELAM}) is proved in the following way.
 By real rescaling, and using the perturbation theory, we obtain,
for fixed $n$, $\,\Omega>0$ and large positive $E,$
\begin{eqnarray}
~~~~~
 \lambda_n(E)\,=
\sqrt{1+2E\Omega}\,\,\bigl(2n+1+O(\Omega/(1+2E\Omega)^{3/2}\,\bigr)\ll E+\Omega E^2.
~~~~~
\label{lambda-E}
\end{eqnarray}
The same behavior applies for  $E>0$ and small $\Omega,$ so that in this regime
the solution of (\ref{RELAM}) is unique.$~~~~~\Box\spazio{0.8}$

Let us now consider how these results can be brought to bear to the study of the
Dirac equation.
From(\ref{psipm}) we see that the components $X_1(x)$ and $X_2(x)$ are linear combinations of two
solutions $\psi_\pm(x)$  of the second order equations. The eigenstates of the
two Titchmarsh operators are independent because they are complex conjugate with non-vanishing real
and imaginary parts, as it can be seen from the equations and their asymptotic behaviors.
If we take a pair of eigenfunctions $(\psi_+(x),\psi_-(x))$ of the separated equations such that $\psi_+(x)=i\,\overline{{\psi}_-(x)}$, this pair gives directly metastable states of the full problem, whereas the resonances give metastable states only by a cut-off: it is relevant that we have more exact informations \cite{BG, SH} on the metastable states than on the resonances. In the non-relativistic limit $\Omega\ra 0$, the energy levels
$E_n(\Omega)$  tend to the eigenvalues
$E_n$ of the Schr\"odinger Hamiltonian and both the states  tend to the real Schr\"odinger eigenstate $\psi_n$ corresponding to $E_n$.
Recalling again (\ref{psipm}) we also remark that, for the metastable states, $X_2\sim {\rm{Re}}\,\psi_+$ and $X_1\sim {\rm{Im}}\, \psi_+$. The sizes of the large and small components of the spinors, therefore, turn out
to be different by many orders of magnitude, providing an approximation to levels of the metastable states which
is very natural and more accurate than the one we could obtain starting from the Schr\"odinger eigenvalues and
perturbing them by means of the relativistic terms: indeed we the metastable levels are approximated by the
DBS that gives the sum of the complete perturbation series in $(1/c^2)\,$.

Take then the real operator
\be
\widetilde{K}({{\Omega}},E)=\widetilde {H}({{\Omega}},E)+2i{\sqrt{\Omega}} x\,.
\label{HKG}
\ee
This is just the Klein-Gordon Hamiltonian with quadratic electrostatic
potential and it is defined as a closed {\it{PT}}-symmetric operator
$K_+({{\Omega}},E)$ by the behavior of the fundamental solutions,
\begin{eqnarray}
\Phi_\pm(x)\sim \frac1{|x|}\,\exp\bigl({iS(x)}\bigl)
\end{eqnarray}
as $x\ra \pm\infty\,$.
$K_+$  has positive eigenvalues $\lambda_n$
 for any real parameter $E$ \cite {BG,SH}.   Again the equivalent adjoint Hamiltonian,
$$K_-=PK_+P=TK_+T=K_+^*$$
is obtained by parity transform.
Following the same proof of Theorem 2, we can show that there are \textit{infinite positive energy levels} as in the Dirac case. The boundary conditions are similar to the resonance ones, but there are always \textit{mixed conditions} Gamow, anti Gamow at $\pm\infty.$
We define as the {\textit{PT}-symmetric pair of dynamics} the dynamics generated by the two Hamiltonians $K_\pm.$ It is interesting to note that the energy levels are the same for the two dynamics, and that the two states are  complex conjugate.
$\spazio{0.8}$

\rmk 21
Let us stress that the small difference between
$\widetilde{H}_+$ and $\widetilde{K}_+$ is, however,
relevant: the formal
operator  $\widetilde {K}$ \textit{is in fact not uniquely implemented
as a closed operator}. In particular, we have infinite self adjoint extensions with  discrete spectrum. Moreover, there is the closed extension  defined by the Gamow condition at $\pm\infty,$  with eigenvalues considered  resonances, in a generalized sense, given by the Distributional Borel sum of the perturbation series.

In perfect analogy with the $H_\pm$ Hamiltonians, we consider  a pair of isospectral $PT$-symmetric Hamiltonians $K_\pm$ with positive simple eigenvalues. In this problem, where \textit{the} physical dynamics does not exists because is not unique, it is appropriate to consider as physical levels, in first approximation, the values taken from the common eigenvalues of the pair of Hamiltonians $K_\pm$. The corresponding eigenstates of the pair of Hamiltonians $K_\pm$ and their combinations, as the real means, are metastable states for all the infinite dynamics of the problem. The real part of the resonances looks less physical of the eigenvalues of $K_\pm$, even if are very similar for small $\Omega$. Let us remark that probably it is not appropriate to impose the same principles of QM to the relativistic QM, where the pair production effect exists. But it is appropriate to pay the minimum price.
$\spazio{1.0}$

For real energy $E$, we have computed the eigenvalues of (\ref{OFKG}) by a  variational
method called the matrix moment
method. Since each eigenvalue $\lambda_n$ is positive,\cite {SH}, we get the corresponding positive
energy level (\ref{RELAM}) as in the Dirac case.

\bigskip

\sect{Definition of the Hamiltonians by subdominant behavior on two disjoint sectors.}
In this section we will give the prescription for a non-ambiguous definition
of the possible operators we
can obtain from the formal operator  $\widetilde {H}({{\Omega}},E)$ with positive
parameters $({\sqrt{\Omega}},E)$. We
follow the methods and the terminology of references \cite {BG,S}.
Let us fix  $\arg(ix)=\phi,$ and consider the six sectors:
$$S_j=\{-{\pi}/{6}<(\phi-j{\pi}/{3})<{\pi}/{6}\},\quad -2\leq j\leq 3.$$
In particular, we define subdominant in the sector $S_j$, $-2\leq j\leq 3,$
 the
solution $\Psi_j$ of the second order equation,
$\widetilde {H}({{\Omega}},E)\Psi_j=0,$
with the principal behavior  for $\phi=j{\pi}/{3},$
\begin{eqnarray}
\ln (\Psi_j(x))=-\sqrt{\Omega}\,\frac{|x|^3}{3}(1+O(|x|^{-2})){\mathrm{as}}\quad |x|\ra\infty.
\end{eqnarray}
In certain cases, the $L^2$ behavior extends partially to the closure $\bar{S}_j$
of the angular sectors.
In particular, the {\it{PT}}-symmetric Hamiltonian ${H}_+({{\Omega}},E)$ is defined by the
subdominant behavior on the
pair of sectors $(S_{-2},S_2).$ Actually, the  solutions respectively subdominant
on the two sectors can
be identified with  the two fundamental solutions at $\pm\infty.$

Let us now examine the complex contours that can be taken according to the
operators we want to define.

\textit{ $(i)$ Complex translation and distortion}.
Let us start again from  $H=\widetilde {H}_+({{\Omega}},E)$ for positive $({\sqrt{\Omega}},E)$,
as above. Consider next
the complex translation $$\psi(x)\mapsto T_z\psi(x)=\psi(x+z)\,,$$
and the translated operator
\begin{eqnarray}
&H^z=T_z\,H\,T_{-z}\,,\spazio{1.}\cr
&{H}^z({{\Omega}},E)=-{\displaystyle{\frac{d^2}{dx^2}}}-{\Omega}E^2-E - 2i{\sqrt{\Omega}} (x+z)
+(1+2E{\Omega})(x+z)^2-{\Omega}(x+z)^4 \,.
\label{Hz}
\end{eqnarray}
For $z=iy$, $y>0$, $H^z$  is uniquely defined as a closed operator
by the fundamental solutions
\begin{eqnarray}
\Psi^z_\pm(x)\sim \frac1{|x|^2}\,\exp\bigl(iS(x+z)\bigr),
\label{yyy}
\end{eqnarray}
as $x\ra\pm\infty,$ respectively. A relevant observation is in order. As the
fundamental solutions are coincident
with the subdominant solutions on the pair of sectors $(S_ {-2},S_2)$,  we can
prove that all the translated
operators are isospectral with the operator defined by the subdominant condition
on the two sectors above.
Indeed it is easy to see that the fundamental solutions of the translated operators
are the translated of
the fundamental solutions of $H_+(\beta,E)$, with the behavior
\be
\Psi_\pm(x)\sim \frac1{|x|^2}\,\exp\bigl(iS(x)\bigl),
%\nonumber
\ee
as $x\ra\pm\infty,$ respectively. Since the eigenvalues are the zeros of the
Wronskian of the two fundamental
solutions, we can conclude that the translated operators are isospectral with
the operator $H_+({{\Omega}},E)$ itself.
In the numerical calculations the translation parameter $y$ will be used as a
variational parameter in
the application of the Rayleigh-Ritz method. We could also consider the
$\mathit{PT}$-symmetry conserving
distortion of entire functions,
$$\psi(x)\mapsto D\psi(x)=\sqrt{z'(x)}\,\,\psi(z(x)),$$
where
$z(x)=x\bigl(1+i\tan({\pi}/{6})\,x\,\sqrt{1+x^2}\,\bigr)$,
\cite {JM}, yielding strictly sectorial operators,
$$H^D_+({{\Omega}},E)=DH_+({{\Omega}},E)D^{-1}.$$
These will be treated elsewhere.

\textit{ $(ii)$ Complex dilation}.
Let  $H=\widetilde  H_+({{\Omega}},E)$ as above, with ${{\Omega}}>0$. Consider first
the complex dilation
\begin{eqnarray}
\psi(x)\mapsto T_\theta\psi(x)=\psi(x\exp(i\theta))=\psi^\theta(x),
\nonumber
\end{eqnarray}
and the operator
\begin{eqnarray}
\widetilde {H}_\theta(\Omega,E)=T_\theta\, \widetilde{H}(\Omega,E)\,T_{-\theta}\,,
\label{Htheta}
\end{eqnarray}
where
$\psi(x)$ is restricted to the
dense set of $L^2$ functions  analytic on an angular sector
$$\left |\frac{{\mathrm{Im}} z}{{\mathrm{Re}} z}\right |<\tan(\theta_0)\qquad {\mathrm{with}} \qquad 0<|\theta|<\theta_0<\pi/6\,.$$
Such operators are defined  by the subdominant
behavior on the two pairs
of sectors $(S_{-2},S_1)$ and $(S_{-1},S_{2})$ and they are isospectral
to the dilated ones with negative
and positive $\theta$ respectively. For positive $\theta$ the fundamental
solutions at $\pm\infty$ have
the  behavior
\be
\ln(\Psi^\theta_\pm(x))=i\,S(x\exp(i\theta))\,+\,O(\ln(|x|)),
\label{fundtheta}
\ee
as $x\ra\pm\infty,$ respectively. For negative $\theta$ the fundamental
solutions at $\pm\infty$ are
\be
\ln(\Psi^\theta_\pm(x))=-i\,S(x\exp(i\theta))\,+\,O(\ln(|x|)).
\label{fundtheta1}
\ee
The two DB complex conjugate sums \cite{CGM1,CGM2,CGM3,CGM4,CGM5} apply to the energy levels of
these operators defined  by
the subdominant behaviors on the two pairs of sectors $(S_{-2},S_1)$ and
$(S_{-1},S_{2}),$ respectively.
The application of the perturbation theory leading to the DB sums has been
thoroughly studied in \cite{GG} by numerical methods, specializing, in
particular, to the calculation of the
pair production rate obtained from the imaginary part of the resonance.
\bigskip

\sect{Results and conclusions.}
\label{numer_results}

In order to compare the spectra of different boundary value problems, instead of integrating the differential equations as in \cite{GS}, we have calculated the eigenvalues using a slight modification of the classical Weinstein-Aronszajn method \cite{WA} obtained from the theory of matrix moments, \cite{GG_Matrix}, joint to the Newton procedure for finding the zeros of a function. In fact, due to the Jacobi form (\ref{MatH}) of the Hamiltonian matrix in the base of the occupation number states, this is equivalent to a Rayleigh-Ritz variational scheme made more efficient through the matrix moments that shorten the computational time and increase the precision of the calculation.

We report, in Figure 1 the lowest energy levels of the Dirac electrostatic oscillator for different values of the
parameter $\Omega$.

%
%%%%%%%%%%%%%%%%%%%%%%%%%%%%%%%%%%%%%%%%%%%%%%%%%%%%%
%
\begin{figure}[ht]
\begin{center}
\includegraphics*[height=4.8 cm,width=7.0 cm]{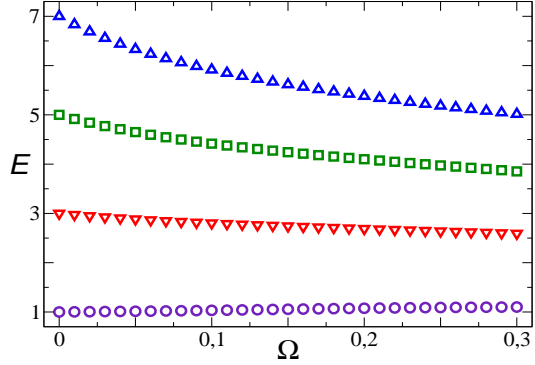}
\caption{The lowest energy levels of the Dirac oscillators for different
values of $\Omega$. Circles refer to the lowest positive energy level. Down triangles, squares and up triangles
refer to the first, second and third excited levels respectively.}
\end{center}
\end{figure}
%
%%%%%%%%%%%%%%%%%%%%%%%%%%%%%%%%%%%%%%%%%%%%%%%%%%%%%
%

We consider the translated, real and dilated operators uniquely defined on
$L^2(\mathbb{R})$. These
are compressed on the space $\mathcal{H}_n$ spanned by the first $n$ Hermite eigenfunctions
$\{\psi_j\}_{j\leq n}$
of the Schr\"odinger Hamiltonian  $H_0=p^2+\sigma^2x^2$, where $\sigma$ is a variational
parameter.
The matrix elements $H_{ik}$ of the fourth degree polynomial in $x$ are obviously dealt with by using the recurrence relation
\begin{eqnarray}
x\,h_k=(1/2)\,h_{k+1} +k\,h_{k-1}
\nonumber
\end{eqnarray}
for the Hermite polynomials $h_r=h_r(x)$. The explicit relevant relations we are using read
\begin{eqnarray}
&{}& x^2\,h_k\! =\! (1/4)\,h_{k+2}\!+\!(k\!+\!1/2)\,h_k+n(k\!-\!1)\,h_{k-2}\spazio{0.8}\cr
&{}& x^3\,h_k \!= \!(1/8)\,h_{k+3}\!+\!(3/4)(k\!+\!1)\,h_{k+1}\!+\!(3/2)\,k^2\, h_{k-1}
+k(k\!-\!1)(k\!-\!2)\,h_{k-3}\spazio{0.6}\cr
&{}& x^4\,h_k = (1/16)\,h_{k+4}+(1/4)(3\!+\!2k)\,h_{k+2}
+(3/4)(2k^2\!+\!2k\!+\!1)\,h_k+k(2k\!-\!1)(k\!-\!1)\,h_{k-2}\spazio{0.6}\cr
&{}& \phantom{x^4\,h_k\!\!\! =\!\!}+k(k\!-\!1)(k\!-\!2)(k\!-\!3)\,h_{k-4}
\nonumber
\label{Matrel}
\end{eqnarray}

\noindent Thus the general form of the matrix reads
\begin{eqnarray}
\bigl(H_{ik}\bigr)=
\left(\begin{array}{ccccc}
A_0 & B_0 & 0 & 0 & \cdots\\
C_0 & A_1 & B_1 & 0 & \cdots \\
0 & C_1 & A_2 & B_2 & 0 \\
\cdots & \cdots & \cdots & \cdots & \cdots
    \end{array}\right)
%\nonumber
\label{MatH}
\end{eqnarray}
where the symbols $A_n$, $B_n$ and $C_n$ represent $4\times 4$ blocks given by
\begin{eqnarray}
&{}& (A_n)_{ik}=H_{4n+i,4n+k} \spazio{0.6}\cr
&{}& (B_n)_{ik}=H_{4n+i,4(n+1)+k}\spazio{0.6}\cr
&{}& (C_n)_{ik}=H_{4(n+1)+i,4n+k}=H_{4n+k,4(n+1)+i} =(B_n)_{ki}
\nonumber
\label{BlockABC}
\end{eqnarray}
with $n=0,1,2,\dots$ and  $i,k=0,1,2,3$.

%%%%%%%%%%%%%%%%%%%%%%%%%%%%%%%%%%%%%%%%%%%%%%%%%%%%%
%
\begin{figure}[ht]
\begin{center}
\includegraphics*[height=4.8 cm,width=7.0 cm]{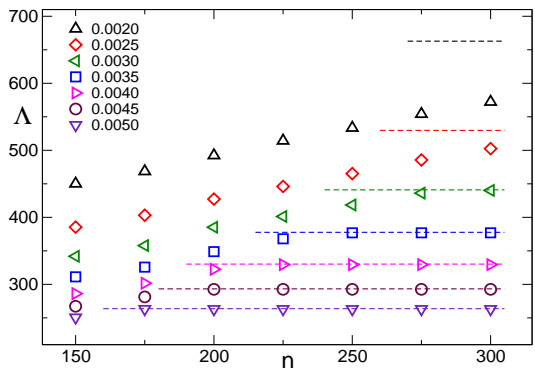}
\caption{The plot of
$\Lambda=-\ln\bigl(E_{t0}-{\rm Re}(E_{d0})\bigr)$ \textit{vs.}
the number $n$ of iterations of (\ref{PolyPn}). Different symbols correspond to different
values of $\Omega$.
The horizontal dashed lines  give $2\,{\rm Im}(E_{d0})$ for the corresponding
$\Omega$.}
\end{center}
\end{figure}
%
%%%%%%%%%%%%%%%%%%%%%%%%%%%%%%%%%%%%%%%%%%%%%%%%%%%%%

The first $4n$ eigenvalues can be approximated by the eigenvalues of the Hamiltonian matrix truncated at the $4n$-th order. From the matrix moment theory \cite{GG_Matrix} these eigenvalues are given by the zeroes of the determinant of the matrix polynomials $P_n=P_n(\lambda)$, recursively defined in the following way:
\begin{eqnarray}
&{}& P_0=I_4\spazio{0.6}\cr
&{}& P_1=B_0^{-1}\bigl(\lambda I_4- A_0\bigr)\spazio{0.6}\cr
&{}& P_n=B_{n-1}^{-1}\bigl(\lambda P_{n-1} - C_{n-2} P_{n-2}- A_{n-1}P_{n-1}\bigr)\,,
%\nonumber
\label{PolyPn}
\end{eqnarray}
$I_4$ being the $4\times 4$ identity matrix.
The solutions of $\det P_n(\lambda)=0$ are indeed the $4n$-th order Rayleigh-Ritz approximants that give upper bounds for the eigenvalues. Although lower bounds for the eigenvalues do not prove to be useful in our present context, it could however be observed that they can be obtained by this same method, looking for the zeroes of the determinant of the matrix polynomial \cite{BF}	
\begin{eqnarray}
P_n(\lambda)-P_n(0)P_{n-1}^{-1}(0)P_{n-1}(\lambda)
\nonumber
\label{PolyBF}
\end{eqnarray}

Let us now present the numerical results. We denote by ``$r,\,t,\,d\,$'' the quantities
respectively related to the spectrum of the real, translated and dilated operators
(\ref{HKG}), (\ref{Hz}) and (\ref{Htheta}) previously discussed. In Table 1,
for different values of $\Omega$, we give the complex
$d$-eigenvalue $E_{d0}$ with lowest
positive real part, reducing to the ground level in the non-relativistic limit.
\begin{table}[hb]
\begin{small}
\begin{center}
\begin{tabular}{c|l}
\texttt{$\Omega$}  & \texttt{\phantom{XXXXXXXXXXXX}$E_{d0}$}\\
\hline
\texttt{0.0020}\, &\,\texttt{1.0005017620\,+\,i\,\,1.17374083059\,e-144}\\
\texttt{0.0025}\, &\,\texttt{1.0006277579\,+\,i\,\,9.42079110945\,e-116}\\
\texttt{0.0030}\, &\,\texttt{1.0007539782\,+\,i\,\,1.72376665081\,e-96}\\
\texttt{0.0035}\, &\,\texttt{1.0008804241\,+\,i\,\,9.77543924661\,e-83}\\
\texttt{0.0040}\, &\,\texttt{1.0010070969\,+\,i\,\,2.00211928567\,e-72}\\
\texttt{0.0045}\, &\,\texttt{1.0011339978\,+\,i\,\,2.08165603853\,e-64}\\
\texttt{0.0050}\, &\,\texttt{1.0012611278\,+\,i\,\,5.36447802132\,e-58}\\
\end{tabular}
\end{center}
\label{tavola1}
\end{small}
\caption{The lowest $d$-eigenvalue $E_{d0}$ for varying $\Omega$.}
\end{table}
Since the effect to be highlighted is really tiny, these eigenvalues have been calculated with great accuracy, both in the arithmetic
precision and in the number of iteration of the recurrence relation (\ref{PolyPn}).
The largest value of the
latter has been taken to be 300: although a  large number of decimal figures is
already stabilized by few iterations, a very high precision is however necessary for a
comparison of the
\textit{t} with the \textit{d} eigenvalues, that we show here below. As expected, the
results
obtained in  \cite{GS,GG} when studying the resonances of the Dirac equation by the
spectral concentration
and by the DB sum, as well as the asymptotic behavior of the imaginary part, are
confirmed by the
much more precise data given in  Table 1. In the application of the Rayleigh-Ritz method we have
used as variational parameters the size of the imaginary translation $y$ introduced in
item (\ref{yyy})
and the the `frequency' $\sigma$ of the operator $H_0$. There has been numerical evidence
that the optimal values
of these parameters are $1\lesssim y\lesssim 5$ and $1 \lesssim \sigma \lesssim 2$.
In Figure 2 we plot
$$\Lambda=-\ln\bigl(E_{t0}-{\rm Re}(E_{d0})\bigr)$$
for a different number $n$ of iterations of (\ref{PolyPn}) in order to test the stabilization
of the data with $n$.
We see that the saturation for decreasing values of $\Omega$  requires  increasing
values of $n$: it appears,
however, that $n=300$ is already sufficient for $\Omega=0.0030$ and higher. The final
value
is almost coincident with minus twice the logarithm of the imaginary part of the eigenvalue
$E_{d}$, represented
by the dashed horizontal lines.
\smallskip

\rmk 41 This means that each energy level is given by the distributional Borel sum of the perturbation series modulo a correction of the second order on the pair production effect (see \cite{GG}).$\spazio{0.8}$

This property has been numerically checked by calculating
the ratio $\Lambda/\bigl(-2\ln{\rm Im}(E_{d0})\bigr)$ for $0.0030\leq\Omega\leq0.0050$.
A simple quartic Lagrangian interpolation on the data thus obtained
gives $0.999885$ as a limiting value of the ratio.

We finally want to consider the influence of the imaginary term proportional to ${\sqrt{\Omega}}$ in
(\ref{Hz}).
We have therefore calculated the difference of the first $t$ and $r$ eigenvalues
and the ratio
$$\kappa=\bigl(E_{t0}({\Omega})-E_{r0}(\Omega)\bigr)/\Omega\,.$$
A least square computation on the data for $0.002<\Omega<0.005$ gives
$\kappa=0.9999539755+0.0284823904\,\Omega\,,\,$
providing a numerical evidence that the difference of the previous eigenvalues
vanishes in the limit of a vanishing $\Omega$.$\spazio{0.8}$

We have compared the differences of the first excited minus the fundamental level
for the $t$ and $r$ operators:
$$
\delta=\bigl[\bigl(E_{t1}(\Omega)-E_{t0}(\Omega)\bigr)-
\bigl(E_{r1}(\Omega)-E_{r0}(\Omega)\bigr)\bigr]/\Omega
.$$
finding a more than linear vanishing  behavior with vanishing $\Omega$.

To conclude, in this paper we have completed the rigorous analysis of the basic model
of the harmonic oscillator in 1D Dirac and Klein-Gordon theory,
using recent mathematical results concerning the anharmonic oscillators
and considering the relevance of the physical symmetries \cite{BG,BB}.
As we said, the result is satisfying,
since  it brings it one step closer to the QFT symmetries.
We have also shown that discrete energy values, the exceptional points, can be obtained
by the approximation of decoupling two equations, but without the presence of a scalar potential, whose physical interpretation
is usually uncertain. The philosophy could be that: in relativistic QM it is not possible or convenient to keep all the principles of QM, but we can try to minimize the effects of the pair production. We have PT-symmetric energy levels comparable to the non relativistic ones. This way, the Klein Gordon equation appears more similar to the Dirac one. In this case, the PT-symmetric eigenvalues appears more physical than the infinite ones  of the infinite selfadjoint Hamiltonians. The levels of the 2D and 3D Dirac oscillators will appear in a following paper.

%============================================================

%======================================================================================

\end{document}